\documentclass{aastex631}
\usepackage{amsmath}

\numberwithin{equation}{section}

\begin{document}

\title{An eccentric transit timing test of modified gravity}

\author{Benjamin Monreal}
\email{benjamin.monreal@case.edu}
\affiliation{Department of Physics, Case Western Reserve University}

\author{Xavier Moskala}
\affiliation{Department of Physics, Case Western Reserve University}

\author{Sofia Splawska}
\altaffiliation{Current address: Department of Physics, Carnegie Mellon University}
\affiliation{Department of Physics, Case Western Reserve University}

\begin{abstract}
  The MOND modified gravity paradigm, best known for its agreement with galactic rotation curve data, is difficult to devise laboratory tests for.   MOND's predictions differ substantially from Newtonian gravity only in the case of very small accelerations ($a < a_0 = 1.2\times10^{-10}~\mathrm{m/s}^2 = 3.8~\mathrm{mm/s/y}$).   In the solar system, radio and laser measurements of test bodies do permit acceleration measurements this precise; however, in at least some viable realizations, MOND has an ``external field effect'' (EFE) which essentially disables the MOND effects in the presence of the inner Milky Way's background gravitational field, invalidating Solar System limits.  Can we do any measurement with Solar-System-like acceleration precision, but on a system many kiloparsecs away that avoids the EFE?  In this paper, we show that MOND's non-$1/r^2$ gravitational acceleration has a unique and hard-to-fake effect: in contrast to the non-precessing elliptical orbits predicted by two-body Newtonian gravity, in the presence of MOND we expect apsidal precession in two-body systems.  An extreme precision apsidal precession detection opportunity is available for eclipsing binaries, where, for some viewing geometries, apsidal precession causes transit time variations (TTVs) and transit duration variations (TDVs).  If appropriate binaries can be found and measured, the presence or absence of timing variations may provide a definitive ``solar-system-like'' test of many EFE-bearing MOND theories like QUMOND. 
\end{abstract}


\section{Introduction}

General Relativity as a theory of gravity has been subjected to hundreds or thousands of precision tests.  Some tests have been motivated by specific non-GR hypotheses, while others have proceeded from a general exploratory desire to test GR in new regimes or at new precision.  The theory of modified newtonian dynamics (MOND) is a modified-gravity hypothesis motivated primarily by dark-matter-free explanations of patterns observed (\cite{McGaugh:2000sr}) in galactic rotation curves.  MOND theories result in non-GR/non-Newtonian gravitational forces in a particular weak-field limit.

MOND theories start with an observation-driven statement: in rotation-supported galaxies, the rotation curve $v(r)$ can often be predicted using the visible matter distribution plus Newtonian gravity, \emph{except} in locations where Newton's Law predicts a low acceleration $a < a_0$ where $a_0 \approx 1.2\times10^{-10}$ m/s$^2$; a striking early observation was that this same constant seems to works over a wide range of galaxy masses.   If you further state that, around $a < a_0$, the gravitational acceleration law ``simply'' crosses over from a $1/r^2$ force to, specifically, a $1/r$ force, the dark-matter-less predictions continue to match galactic observations fairly well. Although similar dynamics can emerge from standard cold dark matter (\cite{Dutton:2019gor}), this emergence is not as precise, quantitative, or overconstrained as one might wish.

Writing a consistent Lagrangian mechanics whose resulting forces are MONDian is not trivial, but several realizations have been devised.   These theories are constructed so as to ensure the $1/r$ behavior in the ``deep MOND'' $a < a_0$ limit and Newtonian behavior in high-acceleration and Solar System contexts, but being usable field theories they can also be used to make predictions at intermediate accelerations. In this paper, we focus on QUMOND (\cite{milgromQuasilinearFormulationMOND2010}), which proposes a new field $\phi$ whose gradient results in gravitational-field-like forces.

QUMOND's dynamics, like those of, e.g., AQUAL (\cite{milgromSolutionsModifiedNewtonian1986}), include a phenomenon called an {\em external field effect} or EFE (\cite{chaeNumericalSolutionsExternal2022}) which is important for MOND testability.  In the Solar System, the Sun's mass (for example) and the Milky Way's mass both contribute to the local solution for $\phi$.  The solution in this case is dominated by Milky Way-sourced terms; the Sun is unable to source a field that contributes a $1/r$ behavior, so Newtonian $1/r^2$ gravity is the only thing observable\footnote{The proposed KBO signal (\cite{brownModifiedNewtonianDynamics2023}) is not a MONDian solar force, but rather a side effect of galaxy-scale fields.}.  Things are different for a star in the outer Milky Way.  There, with the galaxy-sourced $\phi$ field weaker, a star-sourced $\phi$ field can create a local $1/r$ force.  Thus, QUMOND is not ruled out by the nonobservation of $1/r$ forces in the Solar System (\cite{blanchetTestingMONDSolar2011}) since the Sun's galactic environment will have subjected it to the external field effect.  Gravitational tests analogous to solar system tests {\em could} discover or rule out QUMOND, if they could be performed on systems further from the galactic center.

The goal of this paper is to establish (a) that the apsidal precession of an eccentric orbit is sensitive to $1/r$ gravitational force law terms like those of QUMOND, (b) that precision eclipse timing observables are sensitive to apsidal precession, and (c) that eclipse timing of outer-galaxy binary stars and/or planets is a plausible new test of QUMOND.

We focus on QUMOND, rather than other relativistic MOND models, because (despite its complicated Poisson-equation apparatus) it reduces to a straightforward analytic central force law (\cite{zhaoModifiedKeplerLaw2010}) in the two body problem.  We can use this in otherwise-conventional dynamics calculations.  A more complete survey of the MOND model space is beyond the intended scope of this paper.

\section{Transit time and transit duration variations from apsidal precession}

Consider a single body of mass $m$, say a planet\footnote{We're choosing a planet-star system for ease of explanation but will translate these results to binary stars later.}, on an elliptical orbit around a star of mass $M$.  Observing a ``transit'' means, in terms of orbital state vectors, that the planet's true anomaly $\theta(t)$  and the argument of periapsis $\omega$ coincide: $\theta(t) = \omega$.   In the Newtonian two-body problem, the vector orientation of the ellipse (represented by the Laplace-Runge-Lenz vector or the ellipticity vector) is conserved, so the periodically-varying $\theta$ reaches the fixed $\omega$ once per orbit and with a fixed periodicity: $\theta(t_i) -\theta(t_{i+1}) = 2\pi$.

\begin{figure}
  \includegraphics[width=0.4\textwidth]{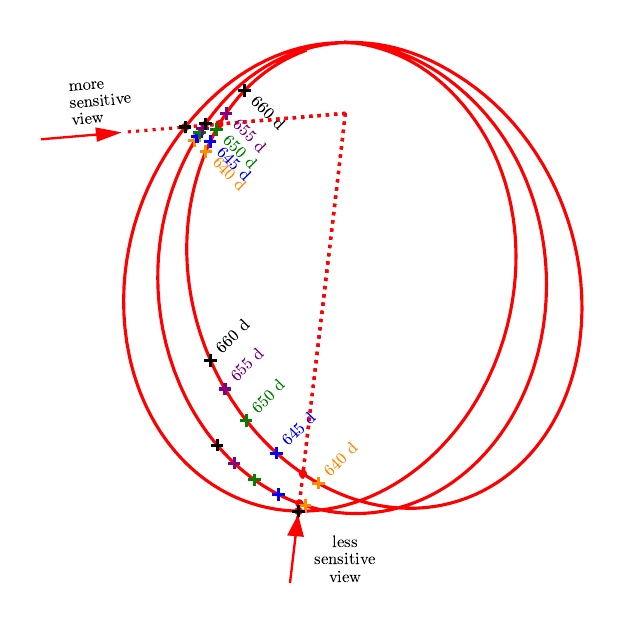}
  \includegraphics[width=0.6\textwidth]{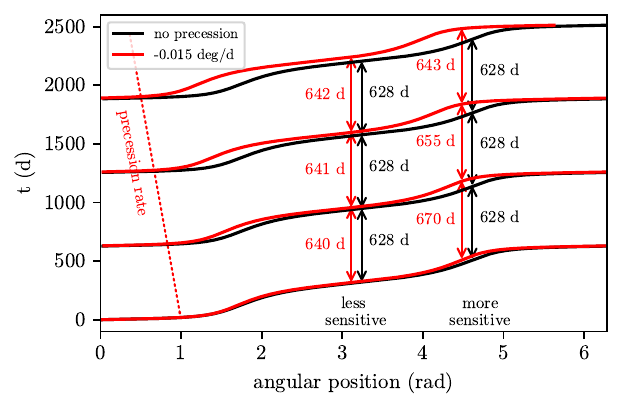}
  \caption{Illustration of the mechanism relating precession to transit time variation.  On the left, we show a pole-down view of one body on a $e=0.7$ orbit with a Keplerian 628-d period and an exaggerated -0.015 degree/day apsidal precession.  An observer positioned in the direction of either arrow would see eclipses when the body crosses the small red dots.  Viewed by the ``less sensitive'' observer near apoapsis, the three eclipses occur at fairly uniform intervals around 641 d, despite taking place at quite different Keplerian phases.  Colored crosses show the positions corresponding to other uniform periods.  The viewer nearer the latus rectum sees three eclipses which do not line up with any constant period; their timing observations are ``more sensitive'' to the precession.  On the right, we show these orbits as orientation vs time, with the precessing system in red and a nonprecessing system for comparison in black.  Eclipses occur when the body returns to the viewer's orientation.  Both observers see a constant 628-day period for the nonprecessing orbit.  For the precessing system, the more-sensitive observer sees eclipses with large timing differences (670, 655, and 643 d) while the less-sensitive observer sees a long (641 d) but almost-constant period.  This illustrates how the transit time variation arises from the second derivative of $t(\theta)$.  Transit durations vary over the course of the orbit, being shorter at periapsis and longer at apoapsis.}\label{fig_illust}
  \end{figure}

Apsidal precession implies a small drift $\omega(t) = \omega_0 + \dot{\omega}t$.   Insofar as the elliptical description is still appropriate, we say that successive transits occur at different true anomalies: $\omega_{n+1} \simeq \omega_n + \dot{\omega}T$ so  $\theta(t_i) -\theta(t_{i+1}) = 2\pi + \dot{\omega}T$.    Due to the nonlinear mapping from $\theta$ to $t$ in elliptical orbits, the addition of {\em constant} increments to the angle can lead to {\em variable} increments in the time.  The principle is illustrated in Fig.~\ref{fig_illust}. (Note that this nonlinearity is absent in circular orbits, and for elliptical orbits it vanishes at $\omega=0$ and $\omega=\pi$, i.e. at apoapse or periapse.)

In the following subsections we will (a) derive the QUMOND prediction for the apsidal precession rate and (b) work through the orbital mechanics to predict the resulting TTVs and TDVs.

\subsection{Apsidal precession in two-body QUMOND}

In a Newtonian $V(r) = -GMm/r$ potential, independent of the mass ratio and the ellipticity, orbits do not precess; the Laplace-Runge-Lenz vector is conserved.  To find the precession rate in the presence of a perturbing potential $h(r)$

\begin{equation}
  V(r) = -\frac{GMm}{r} + h(r)
\end{equation}

is a textbook problem at small eccentricity and a moderately harder one at arbitrary eccentricities; fortunately the literature offers many clear walkthroughs of the solution, e.g. \cite{fitzpatrickIntroductionCelestialMechanics2012}.

Let us find the perturbation $h(r)$ corresponding to the QUMOND force law.  The QUMOND two-body force law is derived in \cite{zhaoModifiedKeplerLaw2010} to be:

\begin{equation}
  F = \frac{GMm}{r^2} + \Xi(M,m)\frac{\sqrt{G(M+m)^3a_0}}{r}
  \end{equation}
  where
  \begin{equation}
    \Xi(M,m) = \frac{2}{3}\left(1 - \left(\frac{M}{M+m}\right)^{3/2} - \left(\frac{m}{M+m}\right)^{3/2}\right)
  \end{equation}
  To describe that force in terms of a potential, we integrate.  Since a constant potential offset has no effect, we do not need to worry about the upper limit of integration or, bizarrely, the choice of $r$ units; the latter is an aspect of the ``scale-free'' nature of MOND (\cite{Milgrom:2008cs}).
  
  \begin{equation}\label{eq_h}
h(r) = -\frac{2}{3}  \left(1 -\left(\frac{m}{m+M}\right)^{3/2}-\left(\frac{M}{m+M}\right)^{3/2}\right) \log (r) \sqrt{a_0 G (m+M)^3}
\end{equation}

The apsidal precession rate for logarithmic perturbations is given analytically in \cite{adkinsOrbitalPrecessionDue2007}.  For the expression in Eq.~\ref{eq_h}, the apsidal precession advance per orbit is


\begin{equation}\label{dTheta_gen}
\delta\theta = \left(\frac{2}{3}(2\pi)^{1/3}\right)\left(\frac{a_0^{1/2}T^{2/3}}{G^{1/6}}\right)\left(\frac{1 - e^2 - \sqrt{1-e^2}}{e^2}\right)\left(\frac{(m^{3/2} + M^{3/2} - (m+M)^{3/2})(m+M)^{1/3}}{mM}\right)
\end{equation}

where $a$ is the semimajor axis, $T$ is the (unperturbed, Keplerian) period, and $e$ is the eccentricity.

\subsection{Transit time sensitivity to apsidal precession}

Next, we can calculate the transit timing effect of apsidal precessions.   In a non-precessing system, every transit occurs at a fixed value of the true anomaly $\theta = 2\pi - \omega$ where $\omega$ is the argument of periapse.  Kepler's Laws relate $t$ and $\theta$ in a way we can solve\footnote{Obtaining $\theta$ given $t$ requires solving a transcendental equation, but this is the simpler inverse problem of finding $t$ given $\theta$.} for $t$:
\begin{equation}
  t= \frac{T}{2\pi} \left(e \sin \left(2 \tan ^{-1}\left(\frac{\sqrt{e-1} \tan \left(\frac{\theta }{2}\right)}{\sqrt{-e-1}}\right)\right)-2 \tan ^{-1}\left(\frac{\sqrt{e-1} \tan \left(\frac{\theta }{2}\right)}{\sqrt{-e-1}}\right)\right)  \label{eq_ttheta}
\end{equation}

In a precessing system, the argument of periapse is advancing by $\delta \theta$ per radial cycle, so the $n$th transit will occur when $\theta_n = 2\pi - \omega_0 - n \delta \theta$ where $\omega_0$ is the initial argument of periapse.   Since $n \delta \theta$ is always small we can can Taylor-expand equation Eq.~\ref{eq_ttheta} to predict the $n$th transit time $t_n$: 

\begin{equation}
  t_n = n\,T + n\,\delta\theta\,\left.\frac{dt(\theta)}{d\theta}\right|_{\theta=\theta_0} + \left.\frac{(n\,\delta\theta)^2}{2}\,\frac{d^2t(\theta)}{d\theta^2}\right|_{\theta=\theta_0} +\;\ldots
\end{equation}

The first term is the radial or Keplerian period of the orbit.  The second term, the first derivative, shifts the transit time interval {\em relative to the Keplerian prediction}.  This would be a new-physics signal only if you knew the precise masses and orbital elements which allowed you to predict the Keplerian period $T$.  However, the interval remains constant for many orbits and would not be observed as a TTV.  The third term, from the second-derivative, contains the TTV we are looking for.

We replace the Keplerian-period term and the first-derivative term with an effective period $T'$.  Carrying out the second derivatives, we find that the transit times now change by $\Delta t$ per orbit, with $\Delta t$ depending on the eccentricity, period, the initial argument of periapse, {\em and} on the MOND-dependent  $\delta\theta$, the apsidal precession angle per orbit.

\begin{multline}
t_n =  nT' - n^2 \Delta t = nT' - n^2 \left[\rule{0cm}{1cm}\left(-\frac{2^{5/3}}{9 \sqrt[3]{\pi }}\right)\left(\frac{a_0 T^{7/3}}{\sqrt[3]{G}}\right) \right.\times  \\*
\left(\frac{\left(1-e^2\right)^{5/2} \left(\sqrt{1-e^2}-1\right)^2 \sin (\omega )}{e^3 (e \cos (\omega )+1)^3}\right) \times \\*
\left.\left(\frac{(m+M)^{11/3} \left(\left(\frac{m}{m+M}\right)^{3/2}+\left(\frac{M}{m+M}\right)^{3/2}-1\right)^2}{m^2 M^2}\right)\rule{0cm}{1cm}\right]
\label{timevar_full}
\end{multline}


We note that the null hypothesis---GR's prediction of zero TTV---is independent of the observer's knowledge (or lack of knowledge) of the masses of the two objects.   MOND's TTV predictions are sensitive to the period (but we measure the period very well) and only weakly sensitive to the object masses.   This is distinctly different from, e.g., statistical wide binary tests of gravity (\cite{scarpaDynamicsWideBinary2017,pittordisWideBinariesGAIA2023,hernandezStatisticalAnalysisGravitational2024}), in which we are comparing observed orbit velocities to modeled ones, and both mass-estimation and radius-estimation uncertainties contribute linearly to the predictions.

The transit intervals may be seen as growing {\em or} shrinking depending on the argument of periapse.  (We presume we will be studying binaries with known arguments of periapse from, e.g., radial velocity data.)

\subsubsection{The test-particle limit and symmetric-binary limits} 

In the test-particle limit of $m << M$ the apsidal precession advance per orbit simplifies to:

\begin{equation}
\delta\theta = \left(\frac{(2\pi)^{1/3}T^{2/3}\sqrt{a_0}}{G^{1/6}}\right)\left(\frac{e^2 + \sqrt{1-e^2} - 1}{e^2}\right)\left(\frac{1}{M^{1/6}}\right)
\label{dtheta_testpart}
\end{equation}

within which the eccentricity-dependence is fairly weak, leaving the useful numerical approximation: 

\begin{equation}
  \delta\theta \simeq 90 \left(\frac{T}{\mathrm{y}}\right)^{2/3} \left(\frac{M}{M_\odot}\right)^{-1/6}~\mathrm{arcsec/orbit}
 \label{dtheta_approx}
\end{equation}

and the full transit time variation simplifies to

\begin{multline}
t_n =   nT' - n^2 \Delta t = nT' - n^2\left[\rule{0cm}{1cm} \left(-\frac{2^{5/3}}{9 \sqrt[3]{\pi }}\right)\left(\frac{a_0 T^{7/3}}{\sqrt[3]{G}}\right) \right. \times \\*
  \left(\frac{\left(1-e^2\right)^{5/2} \left(\sqrt{1-e^2}-1\right)^2 \sin (\omega )}{e^3 (e \cos (\omega )+1)^3}\right) \times \left. \left(\frac{9}{4M^{1/3}}\right)\rule{0cm}{1cm}\right]
\label{timevar_testpart}
\end{multline}


  This also lets us pull out all the unit-bearing quantities and see the source of the TTV:

  \begin{equation}
    \Delta t = 4.0~\mathrm{s}~\left(\frac{a_0}{1.2\times10^{-10} \mathrm{m/s^2}}\right)\left(\frac{T}{\mathrm{y}}\right)^{7/3}\left(\frac{M}{M_\odot}\right)^{-1/3} F(e,\omega)
    \label{eq_dt_simple_tp}
  \end{equation}

  where $F(e,\omega)$ is a dimensionless geometric quantity which for {\em reasonably common} eccentricities and orientations is in the range $-0.5 < F < 0.5$; larger values are possible with both a high eccentricity and a lucky value of $\omega$.

In the symmetric ($m=M$) limit of an equal-mass stellar binary, the prefactor in Eq.~\ref{eq_dt_simple_tp} drops from 4.0~s to 2.0~s.  

\subsection{Transit duration sensitivity to apsidal precession}

The duration of a transit varies in a similar way.  Equation \ref{eq_ttheta} gives the time $t$ as a function of the true anomaly, i.e. the orbit position, $\theta$.  Consider an eclipse centered at $\theta_0$, when the orbit separation is $r_0$, whose first contact is when the bodies' projected separation is $r_c$.  Contact begins at $t(\theta_0 - \frac{r_c}{r_0})$ and ends at $t(\theta_0 + \frac{r_c}{r_0})$, so the eclipse duration for a viewer at $\theta_0$ is

\begin{equation}
  t_d(\theta_0) = 2 \left.\frac{\partial t(\theta)}{\partial{\theta}}\right|_{\theta_0} \frac{r_c}{r_0}
\end{equation}

$t_d$ is sensitive to apsidal precession due to its $\theta_0$ dependence.  

\begin{equation}
  \Delta t_d = \delta\theta \left(\frac{2^{2/3}}{\pi^{1/3}}\right) \left(\frac{e\sqrt{1-e^2} \sin(\omega)}{(1+e \cos(\omega))^2}\right)\left(\frac{r_c T}{(G (m+M) T^2)^{1/3}}\right)
\end{equation}

and plugging in the MOND-predicted $\delta\theta$ yields:

\begin{equation}
   \Delta t_d = \frac{4}{3} r_c T \sqrt{\frac{a_0}{G}} \left(\frac{(m+M)^{3/2}(1 - \left(\frac{m}{m+M}\right)^{3/2} - \left(\frac{M}{m+M}\right)^{3/2})}{mM}\right)  \left( \frac{(e^2-1)(\sqrt(1-e^2)-1)\sin(\omega)}{e(1 + e\cos{\omega})^2} \right) 
\end{equation}

Rearranging unit-bearing quantities, and expressing all masses in units of $M_\odot$,

\begin{multline}
  \Delta t_d = 27.8~\mathrm{s} \left(\frac{r_c}{R_\odot}\right)\left(\frac{T}{\mathrm{y}}\right) \sqrt{\frac{a_0}{1.2\times10^{-10}\mathrm{m/s}}} \left(\frac{(m+M)^{3/2}(1 - \left(\frac{m}{m+M}\right)^{3/2} - \left(\frac{M}{m+M}\right)^{3/2})}{mM}\right)\sqrt{M_\odot} \times \\*
  \left( \frac{(e^2-1)(\sqrt(1-e^2)-1)\sin(\omega)}{e(1 + e\cos{\omega})^2} \right) 
\end{multline}

\subsubsection{The test-particle and symmetric-binary limits}

In the test-particle limit $m<<M$ the TDV per orbit simplifies to

\begin{equation}
  \Delta t_d = 2 r_c T \sqrt{\frac{a_0}{MG}}\left( \frac{(e^2-1)(\sqrt{1-e^2}-1)\sin(\omega)}{e(1+e\cos(\omega))^2}\right)
  \end{equation}

or, consolidating units,
    
\begin{equation}
  \Delta t_d  = 41.8\mathrm{s}\left(\frac{r_c}{R_\odot}\right)\left(\frac{T}{\mathrm{y}}\right) \left(\frac{a_0}{1.2\times10^{-10}\mathrm{m/s}}\right)^{1/2} \left(\frac{M}{M_\odot}\right)^{-1/2} F_d(e,\omega)
  \label{eq_dv_symmetric}
\end{equation}

In the symmetric binary limit $m=M$, in Eq.~\ref{eq_dv_symmetric} the prefactor drops from 41.8~s to 23.1~s.

\section{Can we distinguish MOND-driven TTVs from third-body perturbations?}

The null hypothesis that GR predicts zero TTV is specific to a pure two-body problem.  A multibody system may of course also show TTV, often in geometries where the perturbing planets do not reveal themselves via transits.   When should an observed precession be interpreted as a test of gravity, and when should it be interpreted as due to perturbing bodies?  


For a given system, two-body QUMOND makes a specific narrow prediction; two-body GR makes a specific narrow prediction; three-body effects might be of any magnitude and sign.  Given a large number of test systems, we can imagine three ``simple'' and easily-interpreted outcomes:
\begin{enumerate}
\item Outcome \#1: Several of the systems line up well with the GR prediction; others are scattered widely.  In this case it will be tempting to conclude ``These systems are a mix of two-body and three-body systems, but the two-body systems are GR-like''
\item Outcome \#2: Several of the systems line up well with the QUMOND prediction; others are scattered widely.  In this case it will be tempting to conclude ``These systems are a mix of two-body and three-body systems, but the two-body systems are MOND-like''
\item Outcome \#3: All of the systems have TTVs far from any of the two-body models.  In this case it will be tempting to conclude ``The population of eclipsing binaries systems are not truly two-body systems, and cannot be used in MOND tests.''  
\end{enumerate}

What if many systems show evidence of third-body perturbations but {\em one} lands on either MOND's or GR's two-body prediction?  What if we only find one or two test systems to begin with?  While we can't predict what the data will ultimately reveal, it may be helpful to survey {\em what} TTV amplitudes are expected from third bodies of different sizes on different orbits.

\subsection{Radial velocity and timing signatures of unseen short-period planets} 
Normally, when we discuss the precession of an outer body's orbit due to an inner one, we take the secular approximation---during one orbit of the outer body, the inner body goes through many orbits, giving it the effective gravitational effect of a ring of matter.  The secular precession can be calculated analytically \citep{naozSecularDynamicsHierarchical2013} and is, in the orbits of interest here, smaller than the MOND precession.    However, for the proposed MOND test we are interested in any instantaneous perturbations of the transit times, not in the emergence of genuine apsidal precession from a long-term average.  At any particular time, the inner body may be leading the outer one (and accelerating it) or trailing it (and decelerating it); these phase-dependent accelerations and decelerations can cause transit time variations which are larger than those due to apsidal precession.

While the apsidal-advance-driven TTV or TDV is specific to certain viewing angles, perturbing-body-driven TTV occurs at all viewing angles.  Any ensemble of transiting binary systems will include many ``unlucky orientation'' systems; any TTVs seen in these systems can be used to probe the population statistics of third bodies.  This would inform the interpretation of any TTV seen in the ``lucky orientation'' systems we would consider as MOND probes.  

Can a phase-dependent TTV or TDV fake a MOND signal---or, alternatively, cancel a MOND signal and fake GR?  MOND-driven TTVs have a consistent sign and amplitude.  Phase-dependent TTVs are oscillatory, with a magnitude and sign that varies from one eclipse to the next.  If the perturbing body is not in an orbital resonance, it requires a fairly extreme run of coincidences for the perturbations to line up and fake the MOND signal.  The odds of these coincidences holding goes down with additional eclipse observations.

\subsection{Radial velocity signatures of large outer bodies}
An outer body with a very long orbit may cause what {\em appears} to be a secular apsidal precession if our entire observing epoch is only a fraction of its orbit, as in V889 Aql (\cite{barochAbsoluteDimensionsApsidal2022}) where an inferred 67-year third body has been seen perturbing an 11-day eclipsing binary.

These large planets are responsible for \emph{quite large} radial velocity (RV) amplitudes---hundreds of meters per second---which make it generally likely that the effect is detectable as an RV derivative even over a small fraction of an orbit.  The RV signal may vanish if a particularly-long-orbit planet is caught in its quarter phase, but the precise alignment required is quite rare.   

Triple- and higher multiplicity stellar systems are known to be rarer than binaries; it is not known if the same statistics extend to long-orbit planets.  Whether or not it does, a search focusing on high-eccentricity planets may preferentially pick up the inner components of hierarchical triples, since the Kozai-Lidov mechanism selectively increases the eccentricity of inner planets.  

\subsection{Non-secular timing dynamics for resonant or near-resonant bodies}
Observed exoplanet TTVs and TDVs to date often show extremely large effects associated with near-mean motion resonances \citep{lithwickExtractingPlanetMass2012}.    The orbital perturbations can be quite large, especially when the system is quite close to the mean-motion resonance, since the perturbing object leads (or lags) the transiting star for a long time near conjunction.   While lower-eccentrity orbital TTV and TDVs can be treated analytically \citep{lithwickExtractingPlanetMass2012}, in this work we are interested in higher eccentricities.   The small amplitudes of the MOND-like TTVs and TDVs mean that they can be mimicked by surprisingly small planetesimals.  {\em If} such planetesimals in such resonances are common, then our TTV observable, across all available observed systems, is predicted to be dominated by planet-driven scatter.   This would be true both inside and outside the EFE horizon.  

A perturbing planet can only fake the (secular precession) MOND signal if its effects---which are generally oscillating---line up with the observing epoch in some way that looks monotonic.  (Obviously a system with new-physics sensitivity would be followed up for far longer \citep{miaoTestsConservationLaws2020a}.)  For a near-resonance, the relevant timescale for a monotonicity search is the ``superperiod'' $P_j$

\begin{equation}
  P_j = \frac{1}{|j/T_2 - (j-1)/T_1|}
\end{equation}

which, for the near-$j{:}j{-}1$ resonance of planets at periods $T_1$ and $T_2$, tracks the angles of conjunctions as they rotate around the star.  For a locked resonance, the relevant period is the libration period.  

Although there is a large parameter space worth exploring, in general we find that MOND-like TTVs and TDVs can be mimicked by undetected single second planets, but only if a rare-seeming set of coincidences occur.  A secular-looking TTV or TDV shift can only occur for (a) very light planets (roughly Mercury- to Mars-like) (b) quite close to the mean-motion resonance, and (c) where our data comes from in a narrow window of time when the conjunction is approaching the line of sight.  The population statistics of such planets might be assessed by future transit timing studies and RV studies of systems closer to the Galactic center and where the EFE turns off the MOND precession.



\section{Observational prospects}

\subsection{Observing strategies}
The desired few-second transit timing precision is obviously observationally challenging; this precision will not be available from conventional survey data but will require large-telescope or large-array timing campaigns for extreme timing precision.  Therefore, our observing strategy for locating MOND-testing eccentric binary systems requires at least five transits from first observation to first possible gravity test.   (That is why, unless the observer is very patient, systems with 1--2~y orbits may be more interesting than systems with 4--5~y orbits even though the latter offer higher sensitivity.)  

\begin{enumerate}
\item Single transit discovery.  If existing catalogs don't appear to include the desired targets, we are probably awaiting new discoveries by post-{\em Kepler} surveys.
\item Second transit and period measurement.  Since this transit must be found by a survey, without foreknowledge of the period, we do not expect timing precision better than 10~m.
\item After the 2nd transit, we can plan moderate-precision RV followup to obtain orbital elements and decide if the system is worth targeting for the extreme-timing-precision followup. 
\item Precision transit followup \#1
\item Precision transit followup \#2
\item Precision transit followup \#3
\end{enumerate}
  
Note that the final three measurements need not be {\em consecutive} transits.   When should extreme precision RV observations be made?   Depending on the target list and the available observing resources, this might be an element of target selection before precision timing followups (i.e., use extreme precision RV to pick ``clean'' systems before devoting large-telescope time to a timing campaign) or it might be a post-timing crosscheck (i.e., reserve extreme precision RV instrument time to followup of systems whose TTV or TDVs are already known to be interestingly small.)

\subsection{Catalog targets}
\begin{figure}
  \includegraphics[width=\textwidth]{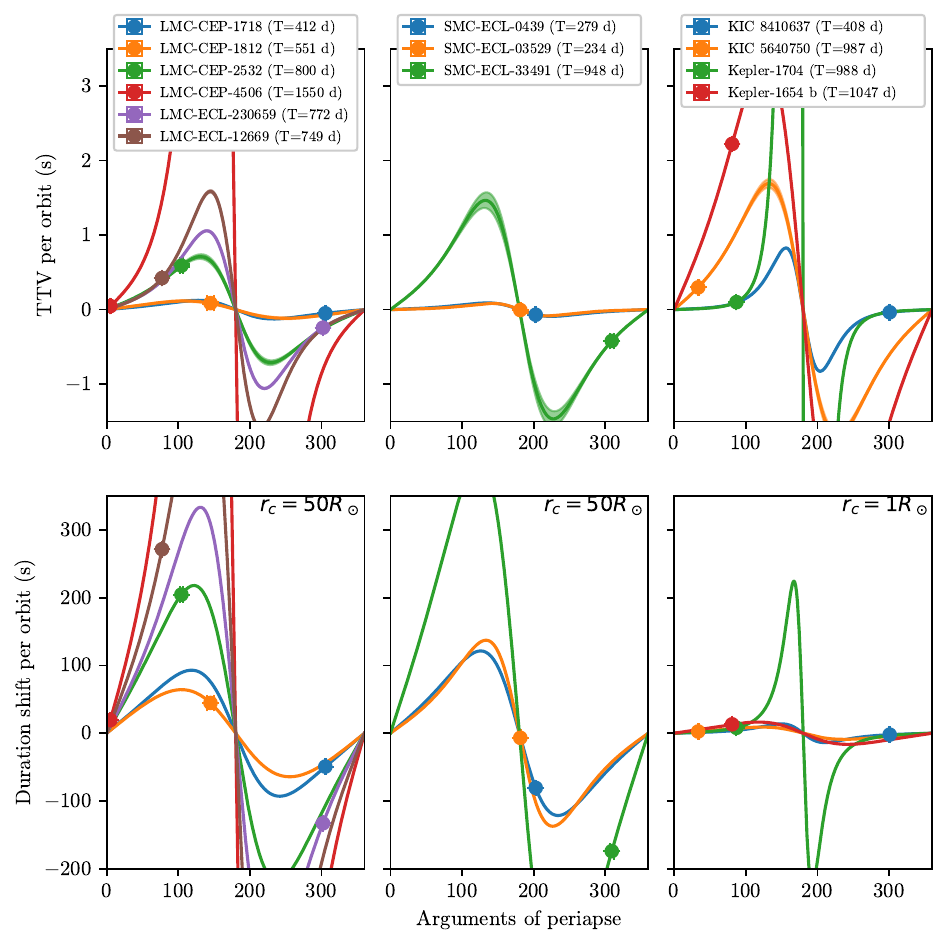}
  \caption{Examples of MOND transit time variation predictions for known long-period eccentric eclipsing systems.  Taking orbit elements from real-world systems, we plot the predicted TTV (top row) or TDV (bottom row, simplified by giving all systems the same eclipse contact radius) as a function of the argument of periapse, as though the system could be viewed from any chosen direction.  (The actual argument of periapse as seen from Earth is marked.)  The top row shows that large transit time variations (multiple seconds) could be possible with ``luckier'' viewing angles which happen not to be found in these samples. The bottom row shows that MOND predicts $\mathcal{O}(100~\mathrm{s})$ transit duration variations in some of these known systems.  Left panel: six eclipsing binary stars found by OGLE in the LMC.  Center panel: three eclipsing binaries found by OGLE in the SMC.  Since the LMC and SMC are weak-potential, EFE-free areas, these systems may be usable as MOND test bodies.  Right panel: long-period, eccentric transiting systems found (in an EFE-affected area) by Kepler, two of which are planets and two of which are binary stars.  In the TDV panel, these four are plotted with the small contact radius more typical of exoplanets, to show how planets' briefer eclipses imply smaller duration shifts.}\label{fig_candidates}
\end{figure}

Are there any candidate transiting systems in present-day catalogs that we might follow up for MOND tests?   For exoplanets, present-day catalogs are dominated by (a) nearby stars observed by TESS, KELT, etc., and (b) more-distant stars in the Kepler fields, which remain deep in the Milky Way.  We do not have examples of transiting exoplanets at the large distances required to escape the EFE.   For eclipsing binary stars, the lengthy OGLE surveys of the Magellanic Clouds discovered numerous examples (\cite{pawlakOGLECollectionVariable2016,graczykLateTypeEclipsing2018,hilditchFortyEclipsingBinaries2005}).  We found several systems for which MOND predicts $\mathcal{O}(1~\mathrm{s})$ variations; however, the periods and eccentricities suggest that $\mathcal{O}(10~\mathrm{s})$ TTV systems will not be uncommon.  MOND TDVs from the same systems may be $\mathcal{O}(100~\mathrm{s})$ or larger.  Fig.~\ref{fig_candidates} shows the sensitivity curves (i.e., the predicted TTVs for alternate imagined viewing angles) for some well-characterized LMC and SMC binary stars.  We also include some Kepler-discovered exoplanets (\cite{td,beichmanValidationInitialCharacterization2018,dalbaGiantOuterTransiting2021}) and binary stars (\cite{gaulmeRedGiantsEclipsing2013}), although we note that the Kepler targets are in EFE-affected locations unsuitable for MOND tests.

However, the ``ideal'' target would be one found roughly towards the Galactic anti-center, where it's possible for a target to be far from the EFE cutoff but close to Earth.  Despite observing difficulties in the center of this field---containing the Milky Way, the ecliptic, {\em and} the Northern Hemisphere---we urge survey planners not to neglect this general direction.

\subsection{Photometry requirements for a MOND search}

\subsubsection{TTV searches}

\begin{figure}
  \includegraphics[width=\textwidth]{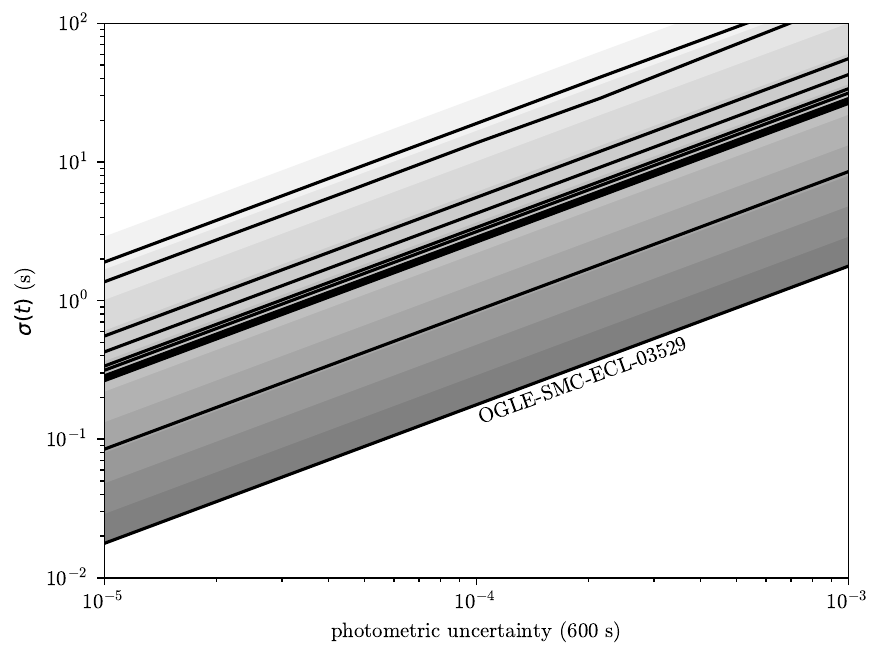}
  \caption{Timing precision vs.\ photometric precision for simulated 600s-cadence observations of plausibly-MOND-sensitive eclipsing binaries.  Using the same OGLE Magellanic Cloud binaries shown in Fig.~\ref{fig_candidates}, we generated model light curves from published orbital elements, using limb darkening coefficients from \cite{1993AJ....106.2096V}.  Gaussian random photometric noise was added, and the curves' time-offsets were scanned with no other free parameters.  Different star systems yielded different levels of timing certainty in a way that largely correlates with eclipse depth.}\label{photo-timing}
\end{figure}

What photometric precision do we need during the ``precision followup'' observations?  This will need to be evaluated separately system-by-system, but for initial guidance we would like to check that MOND-sensitive timing precisions are possible even in idealized data.  For a source of realistic binary star eclipse light curves, we use the same set of OGLE SMC and LMC binaries shown in Fig.~\ref{fig_candidates}.  Using published orbital elements and stellar parameters, we generated simulated light curves and fits assuming 10-minute cadences, and overlaid Gaussian random noise representing a variety of possible photometric noise levels.  We obtain the general photometry/timing trend for wide stellar binaries as shown in Fig.~\ref{photo-timing}.  The results suggest that 0.1~s timing precision often requires space-like $10^{-5}$ photometry.  Although challenging, this is plausibly within reach from the ground, for example by applying the methods in \cite{stefanssonSpacelikePhotometricPrecision2017} to an array of small telescopes.  However, 0.1~s timing of a system like OGLE SMC-ECL-03529 (a particularly deep transit) is possible with $10^{-4}$ photometry.  MOND tests may be possible even with $10^{-3}$ photometry in yet-to-be-discovered systems with higher eccentricities and luckier viewing angles that drive the signal up to 1--10~s.

\subsubsection{TDV searches}

The TDV signature of precession is systematically larger; it might seem like a better observational target than the TTV.  If uncorrelated photometric noise (shot noise, for example) were the main source of error, TDV uncertainties would be of the same order of magnitude as TTV uncertainties.  The problem is: how reliable are transit duration observations?  TTV measurements involve finding the centroid of a dip in the light curve, which can be done reliably even in the presence of various observational uncertainties; as long as any problem affecting the ingress measurement also affects the egress, it doesn't usually shift the centroid.  This is obviously not true of a duration measurement, which might be shifted significantly by observational effects (CCD linearity, for example); by orbital inclination changes; or by stellar activity, including pulsations and limb-darkening variability (\cite{moonVariationSolarLimb2017}).  More study of these systematics is required to understand the prospects for extreme precision in TDV measurements, including in the targets identified above.

\section{Conclusions}
Tests of an EFE-MOND picture of gravity have heretofore mostly been limited to galaxy-scale systems, where its predictions at least substantially overlap with those of cold dark matter cosmology.   There are astronomers who agree, and those who disagree, with propositions that MOND has been successful or unsuccessful in modeling larger-scale systems like the CMB plasma, the Bullet Cluster, large-scale structure, galaxy cluster kinematics, or gravitational lenses, and early structure formation.  However, there is widespread agreement that {\em if} MOND is a force law affecting stars' orbits around galaxies, then it {\em must also} be a force law affecting two-body attraction and motion in some regime, and that a ``clean'' test of that force law would be desirable.  The External Field Effect has prevented such tests from being conducted with the usual laboratory methods. 

In this paper, we have shown how eccentric, transiting binary stars and exoplanets can provide a sensitive test of the two-body force law.  MOND's prediction of a $1/r$ component of this force law leads to a prediction of apsidal precession, even in high-acceleration binary systems where MOND causes a small perturbation.  Apsidal precession can cause transit time variations and transit duration variations, with the largest effects appearing in eccentric orbits with particular viewing angles.  These timing variations can be a high-precision observable in very distant systems, including those where the External Field Effect is weak or absent.  Upcoming time-domain survey projects, including the Nancy Grace Roman telescope, the Rubin Observatory, and PLATO, are expected to find many new transiting binaries, hopefully including some appropriate for high-timing-precision followup.

\section{Acknowledgements}
The authors are grateful for useful conversations with Harsh Mathur, Stacy McGaugh, and Andrius Tamosiunas.  Moskala was supported by the CWRU SOURCE program.  This research has made use of the NASA Exoplanet Archive, which is operated by the California Institute of Technology, under contract with the National Aeronautics and Space Administration under the Exoplanet Exploration Program.

\facilities{Exoplanet Archive, OGLE}

\bibliographystyle{aasjournal}
\bibliography{mond_refs}

\end{document}